\documentclass[11pt]{article}
\usepackage{moriond,epsfig}

\def\bea{\begin{eqnarray}}
\def\eea{\end{eqnarray}}
\def\beq{\begin{equation}}
\def\eeq{\end{equation}}

\newcommand{\ro}{\mbox{{\boldmath
$\rho$}}}

\newcommand{\qb}{\mbox{{\bf
q}}}
\newcommand{\pb}{\mbox{{\bf
p}}}

\newcommand{\Hb}{\mbox{{\bf
H}}}
\newcommand{\Fb}{\mbox{{\bf
F}}}
\newcommand{\Gb}{\mbox{{\bf
G}}}

\newcommand{\mqb}{\bar{\mbox{{\bf
q}}}}

\newcommand{\fb}{\mbox{{\bf
f}}}

\def\lsim{\mathrel{\rlap{\lower4pt\hbox{\hskip1pt$\sim$}}
    \raise1pt\hbox{$<$}}}         
\def\gsim{\mathrel{\rlap{\lower4pt\hbox{\hskip1pt$\sim$}}
    \raise1pt\hbox{$>$}}}         

\begin{document}
\vspace*{4cm}
\title{
SYNCHROTRON-LIKE GLUON EMISSION IN THE QUARK-GLUON PLASMA
}

\author{B.G. ZAKHAROV }

\address{L.D. Landau Institute for Theoretical Physics,
        GSP-1, 117940,\\ Kosygina Str. 2, 117334 Moscow, Russia}

\maketitle\abstracts{
A quasiclassical theory of the synchrotron-like gluon emission 
is discussed. 
We show that the synchrotron emission may be important in 
the jet quenching if the plasma 
instabilities generate a chromomagnetic field $H\sim m^{2}_{D}/g$.
Our gluon spectrum disagrees with that obtained by Shuryak and Zahed
within Schwinger's proper time method.}


\noindent{\bf 1.}
In this talk, I discuss a quasiclassical approach to 
the synchrotron-like gluon emission in the quark-gluon
plasma (QGP) developed in my recent paper \cite{paper}. 
The synchrotron energy loss may potentially be important
in the jet quenching since the plasma instabilities can generate  
chromomagnetic/electric fields in the QGP created in $AA$-collisions 
\cite{M1,Arnold1,Arnold2,M2,Rebhan1}. 
The non-Abelian synchrotron radiation was previously addressed
in the soft gluon limit by Shuryak and Zahed \cite{SZ} within  
Schwinger's proper time method. Our
quasiclassical approach 
is very simple
and applicable in the quantum regime. The gluon 
spectrum derived in our formalism 
disagrees with that obtained in \cite{SZ}. We give arguments that
the spectrum of \cite{SZ} is wrong.
 
\vspace{0.12cm}
\noindent{\bf 2}.
One can show that 
in the quasiclassical regime (when the parton energies are much bigger than 
their masses), similarly to the photon radiation
in QED, the coherence length of the gluon emission, $L_{c}$, is small 
compared to the minimal parton Larmor radius $R_{L}$. 
It allows one to calculate the radiation rate per unit 
length by considering a slab of chromomagnetic field of the 
thickness $L\gg L_{c}$, which is, however, small compared to 
the parton Larmor radii.
In this case the transverse momenta of the final partons
are small compared to their longitudinal momenta (we choose the $z$-axis 
along the initial parton momentum, which is 
perpendicular to the slab with transverse chromomagnetic field, $\Hb_{a}$).
One may consider the magnetic field with the only nonzero 
color components in the Cartan subalgebra, i.e., for $a=3$ and $a=8$ for 
the $SU(3)$ color group. The interaction of the gluons with the external
field is diagonalized by introducing the fields with definite
color isospin, $Q_{A}$, and color hypercharge, $Q_{B}$, (we denote 
the color charge by the two-dimensional vector $Q=(Q_A,Q_B)$).
There are two neutral gluons $A=G_{3}$ and $B=G_{8}$,
and six charged gluons $X,\,Y,\,Z,\,\bar{X},\,\bar{Y},\,\bar{Z}$, where
in terms of the usual gluon vector potential, $G$, 
(the Lorentz indices are omitted)
$X=(G_{1}+iG_{2})/\sqrt{2}$ ($Q=(-1,0)$),
$Y=(G_{4}+iG_{5})/\sqrt{2}$ ($Q=(-1/2,-\sqrt{3}/2)$),
$Z=(G_{6}+iG_{7})/\sqrt{2}$ ($Q=(1/2,-\sqrt{3}/2)$).

The $S$-matrix element of the $q\rightarrow gq'$ transition
can be written as (we omit the color factors and indices)
\beq
\langle gq'|\hat{S}|q\rangle=-ig\int\! dy 
\bar{\psi}_{q'}(y)\gamma^{\mu}G_{\mu}^{*}(y)\psi_{q}(y)\,,
\label{eq:10}
\eeq
where $\psi_{q,q'}$ are the wave 
functions of the initial quark and final quark, $G$ is the wave function
of the emitted gluon.
We write each quark wave function in the form
$
\psi_{i}(y)=\exp[-iE_{i}(t-z)]
\hat{u}_{\lambda}
\phi_{i}(z,\ro)/\sqrt{2E_{i}}$ (hereafter the bold vectors denote the
transverse
vectors), where $\lambda$ is the quark helicity,
$\hat{u}_{\lambda}$ is the Dirac spinor operator.
The $z$-dependence of the transverse wave functions
$\phi_{i}$ is governed by the two-dimensional 
Schr\"odinger equation 
\beq
i\frac{\partial\phi_{i}(z,\ro)}{\partial
z}={\Big\{}\frac{
(\pb-gQ_{n}\Gb_{n})^{2}
+m^{2}_{q}} {2E_{i}}
+gQ_{n}(G^{0}_{n}-G^{3}_{n}){\Big\}}
\phi_{i}(z,\ro)\,,
\label{eq:20}
\eeq
where now $G$ denotes
the external vector potential (the superscripts are the Lorentz indexes and 
$n=A,B$), $Q_{n}$ is the quark color charge. 
The wave function of the emitted
gluon can be represented in a similar way.
We will assume that in the QGP for the parton masses 
one can use the corresponding quasiparticle masses.

We take the external vector potential in the form 
$G^{3}_{n}=[\Hb_{n}\times \ro]^{3}$,
$\Gb_{n}=0$, $G^{0}_{n}=0$ (we assume that chromoelectric field is absent, 
however, it can be included as well). The term $-gQ_{n}G^{3}_{n}$ in
(\ref{eq:20}) can be viewed as a potential energy  $U_{i}=-\Fb_{i}\cdot\ro$,
where $\Fb_{i}$ is the corresponding Lorentz force. 
The solution of (\ref{eq:20}) can be taken in the form
\beq
\phi_{i}(z,\ro)=\exp{\left\{i\pb_{i}(z)\ro-
\frac{i}{2E_{i}}\int_{0}^{z}dz'[\pb^{2}_{i}(z')+m^{2}_{q}]\right\}}\,.
\label{eq:30}
\eeq
Here $\pb_{i}(z)$ is the solution to the equation 
$ 
{d\pb_{i}}/{dz}=\Fb_{i}(z)\,.
$ 
%
From (\ref{eq:10}), (\ref{eq:30}) one can obtain 
\bea
\langle gq'|\hat{S}|q\rangle=-ig
(2\pi)^{3}\delta(E_{g}+E_{q'}-E_{q})
\int_{-\infty}^{\infty}dz V(z,\{\lambda\})
\delta(\pb_{g}(z)+\pb_{q'}(z)-\pb_{q}(z))\nonumber\\
\times
\exp{\left\{-i\int_{0}^{z} dz'
\left[\frac{\pb^{2}_{q}(z')+m^{2}_{q}}{2E_{q}}
-\frac{\pb^{2}_{g}(z')+m^{2}_{g}}{2E_{g}}
-\frac{\pb^{2}_{q'}(z')+m^{2}_{q}}{2E_{q'}}
\right]\right\}}\,,
\label{eq:50}
\eea
where $V$ is the spin vertex factor, $\{\lambda\}$ is the set of the parton
helicities. 
The argument of the transverse momentum $\delta$-function 
does not depend on $z$
(since $\Fb_q=\Fb_g+\Fb_{q'}$),
and can be replaced by $\pb_{g}(\infty)+\pb_{q'}(\infty)-\pb_{q}(\infty)$. 
From (\ref{eq:50}) one can obtain for the gluon spectrum
\bea
\frac{dP}{dx}=
\frac{1}{(2\pi)^{2}}
\int d\pb_{g}(\infty)
\int
dz_{1}dz_{2}
g(z_{1},z_{2})\nonumber\\
\times
\exp{\left\{i\int_{z_{1}}^{z_{2}} dz
\left[\frac{\pb^{2}_{q}(z)+m^{2}_{q}}{2E_{q}}
-\frac{\pb^{2}_{g}(z)+m^{2}_{g}}{2E_{g}}
-\frac{\pb^{2}_{q'}(z)+m^{2}_{q}}{2E_{q'}}
\right]\right\}}\,,
\label{eq:60}
\eea
\beq
g(z_{1},z_{2})=\frac{C\alpha_{s}}{8E_{q}^{2}x(1-x)}\sum_{\{\lambda\}}
V^{*}(z_2,\{\lambda\})V(z_1,\{\lambda\})=
g_{1}\qb(z_2)\qb(z_1)/\mu^{2}+g_{2}\,,
\label{eq:70}
\eeq
where
$x$ is the longitudinal 
gluon fractional momentum,
$\mu=E_{q}x(1-x)$,
$\qb(z)=\pb_{g}(z)(1-x)-\pb_{q'}(z)x$, 
$g_{1}=C\alpha_{s}(1-x+x^{2}/2)/x$ and
$g_{2}=C\alpha_{s}m_{q}^{2}x^{3}/2\mu^{2}$ 
(the two terms in (\ref{eq:70}) correspond to the
non-flip and spin-flip processes), 
$C=|\lambda_{fi}^{a}\chi_{a}^{*}/2|^{2}$, where $i,f$ are the color 
indexes of the initial and final quarks, $\chi_{a}$ is the color wave 
function of the emitted gluon.

For a uniform external field we have
$\qb(z_2)\qb(z_1)=\mqb^{2}-\fb^{\,2}\tau^{2}/4$, where
$\mqb=\qb(\bar{z})$, $\bar{z}=(z_{1}+z_{2})/2)$, $\tau=z_{2}-z_{1}$,
and $\fb=d\qb/dz=\Fb_{g}(1-x)-\Fb_{q'}x$.
After replacing in (\ref{eq:60}) the 
integration over $\pb_{g}(\infty)$ by the integration over $\bar{\qb}$
we obtain for the radiation rate per unit length
\bea
\frac{dP}{dLdx}=
\frac{1}{(2\pi)^{2}}
\int d\mqb
\int_{-\infty}^{\infty} d\tau
\left[\frac{g_{1}}{\mu^{2}}
\left(\mqb^{2}-\frac{\fb^{\,2}\tau^{2}}{4}\right)
+g_{2}\right]
\exp{\left\{-i\left[
\frac{(\epsilon^{2}+\mqb^{2})\tau}{2\mu}
+\frac{\fb^{\,2}\tau^{3}}{24\mu}\right]
\right\}}\,
\label{eq:90}
\eea
with $\epsilon^{2}=m_{q}^{2}x^{2}+m_{g}^{2}(1-x)$.
With the help of $\tau$ integration by parts one can remove $\mqb^{2}$
from the left square brackets in (\ref{eq:90}), and 
after integrating over $\mqb$ one obtains
\beq
\frac{dP}{dLdx}=
\frac{i\mu}{2\pi}
\int_{-\infty}^{\infty} \frac{d\tau}{\tau}
\left[\frac{g_{1}}{\mu^{2}}\left(\epsilon^{2}+\frac{\fb^{\,2}\tau^{2}}{2}\right)
-g_{2}\right]
\exp{\left\{-i\left[
\frac{\epsilon^{2}\tau}{2\mu}
+\frac{\fb^{\,2}\tau^{3}}{24\mu}\right]\right\}}\,.
\label{eq:110}
\eeq
Here it is assumed that $\tau$ has a 
small negative imaginary part.  One can easily show that in (\ref{eq:110}) 
the integral around the lower semicircle near the pole at $\tau=0$ plays 
the role of the $\fb=0$ subtraction term. 
The formula (\ref{eq:110}) can be written in terms of the Airy function
\beq
\frac{dP}{dLdx}=
\frac{a}{\kappa}\mbox{Ai}^{'}(\kappa)+
b\int_{\kappa}^{\infty}dy\mbox{Ai}(y)\,,
\label{eq:120}
\eeq
where 
$a=-{2\epsilon^{2}g_{1}}/{\mu}$,
$
b=\mu  g_{2}-{\epsilon^{2}g_{1}}/{\mu}
$,
$\kappa=\epsilon^{2}/(\mu^{2}\fb^{\,2})^{1/3}$.

\vspace{0.12cm}
\noindent{\bf 3.}
For neutral gluons at $m_{g}=0$ our spectrum (\ref{eq:110}) agrees with 
prediction of the quasiclassical operator approach
\cite{BK} for the photon spectrum.
For charged gluons (\ref{eq:110}) disagrees with the spectrum obtained 
by Shuryak and Zahed \cite{SZ} in the soft gluon limit within Schwinger's
proper time method. In \cite{SZ} 
(Eq. (20) of \cite{SZ}) the argument of the exponential  contains 
(we use our notation) 
$\Fb_{q'}^{2}x_{g}^{2}+\Fb_{g}^{2}$
instead of our 
$\fb^{\,2}$. Also, in the
pre-exponential factor instead of $\fb^{\,2}$ there appears 
$\Fb_{q'}^{2}x_{g}^{2}$.  
Due to the absence of the interference term the spectrum of \cite{SZ} 
is insensitive to the relation between the signs of the color charges 
of the final partons. It is strange enough,
since the difference
in the bending of the final parton trajectories
(which is responsible for the synchrotron radiation) is sensitive
to the relation between the color charges of the final partons.
Also, Eq. (20) of \cite{SZ} in 
the massless limit gives a vanishing spectrum
for the $q_{1}\rightarrow g_{Z}q_{3}$ transition
for magnetic field in the color state $A$  (since in this case $\Fb_{q'}=0$). 
This process except for the spin effects is analogous to the 
synchrotron radiation in QED, and there is no physical reason 
why it should vanish.
One more objection to the result of \cite{SZ} 
is that due to a non-symmetric form of the 
pre-exponential factor in the case of $g\rightarrow gg$ process
it should give  the spectrum with
incorrect permutation properties.
Thus, one sees that the formula obtained in \cite{SZ} clearly
leads to absurd predictions, and cannot be correct.

\vspace{0.12cm}
\noindent{\bf 4.}
In Fig.~1 we present the averaged over 
the color states gluon spectrum for the 
chromomagnetic field in the color state $A$ for different initial parton 
energies. The computations are performed for $\alpha_{s}=0.3$,
and
$gH_{A}/m_{D}^{2}=0.05$, 0.25 and 1, 
where $m_{D}$ is the Debye mass
(we assume that as for an isotropic
weakly coupled plasma $m_{D}^{2}=2m_{g}^{2}$). For the quasiparticle masses
we take 
$m_{q}\approx 0.3$ and $m_{g}\approx 0.4$ GeV \cite{LH}.
For the magnetic field
in the color state $B$ the spectrum is very close to that shown in Fig.~1.
The decrease of the spectra at $x\rightarrow 0$ (and $x\rightarrow 1$ for
$g\rightarrow gg$ process) which is well seen for the smallest value of the
field is due to the Ter-Mikaelian mass effect. This suppression decreases 
with increase of the chromomagnetic field.
%
\begin{figure} [h]
\begin{center}
\hspace{-1cm}\epsfig{file=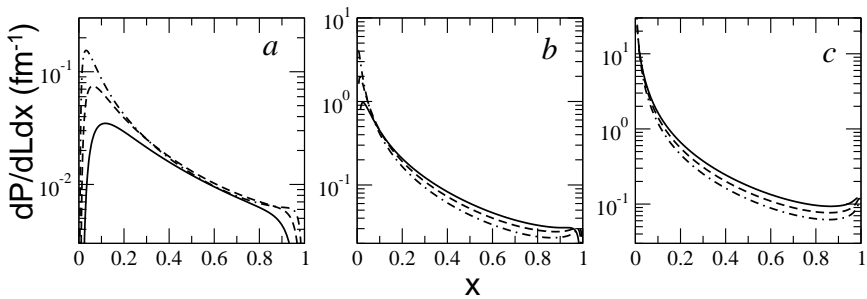,height=10cm,angle=0}
\end{center}
\vspace{-4cm}
\caption[.]
{
The spectrum for the $q\rightarrow gq$ process in the 
chromomagnetic field in the color state $A$ for $\alpha_{s}=0.3$,
$gH_{A}/m_{D}^{2}=0.05$ (a), 0.25 (b) and 1 (c), for the initial quark
energies $E_{q}=20$ GeV (solid line),
$E_{q}=40$ GeV (dashed line),  
$E_{q}=80$ GeV (dash-dotted line).
}
\end{figure}

To estimate the synchrotron energy loss in the QGP produced in $AA$-collisions
we 
take $gH\approx m_{D}^{2}$.
Approximately such a chromomagnetic field is necessary in the scenario
with turbulent magnetic fields \cite{BMueller1} for agreement with 
the small viscosity of the QGP observed
at RHIC.
In this case 
the ratio of the magnetic energy to the thermal parton energy is $\sim 0.3$,
and a higher fraction of the magnetic energy looks unrealistic.
For $\alpha_{s}=0.3$ 
and $L\sim 2-4$ fm
we obtained $\Delta E/E\sim 0.1-0.2$
for quarks and $\Delta E/E\sim 0.2-0.4$ for gluons at $E\sim 10-20$ GeV
(for $\alpha_{s}=0.5$ the results
are about two times bigger). 
However, these estimates neglect any finite-size 
effects which may be important if $L_{c}\gsim L$ \cite{Z_OA}.
The dominating contribution
to the energy loss comes from the soft gluon emission 
where $L_{c}\sim 1-2$ fm. In this situation the finite-size effects
may suppress the energy loss by a factor $\sim 0.5$.
The finite coherence length of the
turbulent magnetic field, $L_{m}$, can suppress the radiation as well. 
If for the unstable magnetic field modes the wave vector 
$k^{2}\lsim \xi m_{D}^{2}$ \cite{BMueller1} 
($\xi\sim 1$ is the anisotropy parameter), 
this suppression should not be very strong since
we have $L_{m}/L_c\sim 1$.
As a plausible estimate
one can take the turbulent suppression factor $\sim 0.5$. Even with
these suppression factors the synchrotron loss 
turns out to be comparable with  the collisional energy loss 
\cite{Z2007,Z2008}, 
which in turn is about 20-30\% of the radiative energy loss. 
Thus our analysis demonstrates that the synchrotron radiation 
can be important in jet 
quenching and deserves further more accurate investigations.
In particular, it would be interesting to treat the synchrotron radiation
and emission due to multiple rescatterings on even footing.
This can be done within the light-cone path integral formalism
\cite{Z1}.

\vspace{.12cm}
\noindent {\bf 5}. 
In summary, we have developed a quasiclassical theory of the 
synchrotron-like gluon radiation. 
Our calculations show that the parton energy loss due to
the synchrotron radiation may be important in the jet quenching
if the QGP instabilities generate magnetic field
$H \sim m_{D}^{2}/g$.
Our gluon spectrum disagrees with that
obtained by Shuryak and Zahed \cite{SZ}.
We give simple physical arguments that the spectrum derived in \cite{SZ} 
is incorrect.

\section*{Acknowledgments}
I am grateful to the organizers for such an enjoyable and stimulating meeting
and for financial support of my participation.

\end{document}